\documentclass[12pt]{iopart}

\usepackage{graphicx}
\usepackage{dcolumn}
\usepackage{bm}

\newcommand{\bq}{\begin{equation}}
\newcommand{\eq}{\end{equation}}
\newcommand{\bqa}{\begin{eqnarray}}
\newcommand{\eqa}{\end{eqnarray}}
\newcommand{\nn}{\nonumber \\}

\def\be     {\begin{equation}}
\def\ee     {\end{equation}}
\def\bnn    {\begin{eqnarray*}}
\def\enn    {\end{eqnarray*}}

\begin{document}

\title{Anomalous transport phenomena in Weyl metal beyond the Drude model for Landau's Fermi liquids}
\author{Ki-Seok Kim$^{1,2}$, Heon-Jung Kim$^{3}$, M. Sasaki$^{4}$, J.-F. Wang$^{5}$, and L. Li$^{5}$}
\address{$^{1}$Department of Physics, POSTECH, Pohang, Gyeongbuk 790-784, Korea \\ $^{2}$Institute of Edge of
Theoretical Science (IES), Hogil Kim Memorial building 5th floor, POSTECH, Pohang, Gyeongbuk 790-784, Korea \\
$^{3}$Department of Physics, College of Natural Science, Daegu University, Gyeongbuk 712-714, Korea \\
$^{4}$Department of Physics, Faculty of Science, Yamagata University, Kojirakawa, Yamagata 990-8560, Japan \\
$^{5}$Wuhan National High Magnetic Field Center (WHMFC), Huazhong University of Science and Technology, Wuhan 430074, China}
\ead{$^{1,2}$tkfkd@postech.ac.kr}
%\ead{$^{1,2}$tkfkd@postech.ac.kr, $^{3}$hjkim76@daegu.ac.kr, and $^{4}$sasaki@sci.kj.yamagata-u.ac.jp}
\date{\today}

\begin{abstract}
Landau's Fermi-liquid theory is the standard model for metals, characterized by the existence of electron quasiparticles near a Fermi surface as long as Landau's interaction parameters lie below critical values for instabilities. Recently, this fundamental paradigm has been challenged by physics of strong spin-orbit coupling although the concept of electron quasiparticles remains valid near the Fermi surface, where the Landau's Fermi-liquid theory fails to describe electromagnetic properties of this novel metallic state, referred to as Weyl metal. A novel ingredient is that such a Fermi surface encloses a Weyl point with definite chirality, referred to as a chiral Fermi surface, which can arise from breaking of either time reversal or inversion symmetry in systems with strong spin-orbit coupling, responsible for both Berry curvature and chiral anomaly. As a result, electromagnetic properties of the Weyl metallic state are described not by conventional Maxwell equations but by axion electrodynamics, where Maxwell equations are modified with a topological-in-origin spatially modulated $\theta(\bm{r}) \bm{E} \cdot \bm{B}$ term. 
%
%In this respect the Weyl metallic state may be regarded as a novel metallic phase, which should be distinguished from the Landau's Fermi-liquid fixed point described by Landau's Fermi-liquid theory.
%
%The Boltzmann transport theory with the Drude model is our standard framework for metals described by Landau's Fermi-liquid theory. However, recently observed anomalous transport phenomena imply that the Drude model has to be modified even if electron correlations are weak and the concept of electron quasiparticles is valid near a Fermi surface. A novel ingredient turns out to be that such a Fermi surface encloses a Weyl point with definite chirality, referred to as a chiral Fermi surface, which can arise from breaking of either time reversal or inversion symmetry in systems with strong spin-orbit coupling.
%
This novel metallic state has been realized recently in Bi$_{1-x}$Sb$_{x}$ around $x \sim 3\%$ under magnetic fields, where the Dirac spectrum appears around the critical point between the normal semiconducting ($x < 3\%$) and topological semiconducting phases ($x > 3\%$) and the time reversal symmetry breaking perturbation causes the Dirac point to split into a pair of Weyl points along the direction of the applied magnetic field for such a strong spin-orbit coupled system.
%
%As a result, this metallic state consists of a pair of chiral Fermi surfaces which encloses each Weyl point, referred to as Weyl metal, where not only the Berry curvature but also the chiral anomaly between the pair of chiral Fermi surfaces is responsible for anomalous transport phenomena such as anomalous Hall effect, chiral magnetic effect, negative longitudinal magneto-electrical resistivity, and so on beyond the description of the Drude model.
%
In this review article, we discuss how the topological structure of both the Berry curvature and chiral anomaly (axion electrodynamics) gives rise to anomalous transport phenomena in Bi$_{1-x}$Sb$_{x}$ around $x \sim 3\%$ under magnetic fields, modifying the Drude model of Landau's Fermi liquids.
\end{abstract}

\maketitle

\section{Introduction}

The Drude model is the first theory that we learn even in the graduate course of the solid state physics \cite{Solid_State_Textbook}. Although we do not have any idea when we learn the theory first, we become surprised later at the fact that such a simple point of view can be realized in much complex organized structures of solids. Learning the Bloch's theorem \cite{Solid_State_Textbook}, we start to understand why essentially the free electron theory works. However, this is just the beginning of a series of surprises in realizing the power of the Drude model because the Drude model still works so well even if electron correlations become strong enough to enhance the band mass of an electron much larger than that of weakly correlated metals. This deep question is resolved within Landau's Fermi-liquid theory, the backbone of which is the existence of electron quasiparticles, where the strongly correlated metallic state is connected to a degenerate Fermi-gas state adiabatically even if the weight of the energy state and the effective band mass near the Fermi surface become renormalized rather much \cite{LFLT_Textbook}.

Recently, this common belief has been challenged by the discovery of a novel metallic state, referred to as Weyl metal \cite{Haldane,Weyl_Metal1,Weyl_Metal2}. Even if the concept of electron quasiparticles remains valid, Landau's Fermi-liquid theory fails to explain anomalous transport properties of this new metal phase. First of all, the characteristic feature of Weyl metal originates from its band structure. Suppose the band structure of a topological insulator, described by an effective Dirac Hamiltonian in momentum space \cite{TI_Band_Structure}
\bqa && H_{eff} = \int \frac{d^{3} \boldsymbol{k}}{(2\pi)^{3}} \psi_{\sigma\tau}^{\dagger}(\boldsymbol{k}) \Bigl( v \boldsymbol{k} \cdot \boldsymbol{\sigma}_{\sigma\sigma'} \otimes \boldsymbol{\tau}_{\tau\tau'}^{z} + m(|\boldsymbol{k}|) \boldsymbol{I}_{\sigma\sigma'} \otimes \boldsymbol{\tau}_{\tau\tau'}^{x} - \mu \boldsymbol{I}_{\sigma\sigma'} \otimes \boldsymbol{I}_{\tau\tau'} \Bigr) \psi_{\sigma'\tau'}(\boldsymbol{k}) . \nonumber \eqa
%
%\bqa && Z = \int D \psi_{\sigma\tau}(\boldsymbol{k}) \exp \Bigl\{ - \int_{0}^{\beta} d \tau \int \frac{d^{3} \boldsymbol{k}}{(2\pi)^{3}} \psi_{\sigma\tau}^{\dagger}(\boldsymbol{k}) \Bigl( (\partial_{\tau} - \mu) \boldsymbol{I}_{\sigma\sigma'} \otimes \boldsymbol{I}_{\tau\tau'} \nn && + v \boldsymbol{k} \cdot \boldsymbol{\sigma}_{\sigma\sigma'} \otimes \boldsymbol{\tau}_{\tau\tau'}^{z} + m(|\boldsymbol{k}|) \boldsymbol{I}_{\sigma\sigma'} \otimes \boldsymbol{\tau}_{\tau\tau'}^{x} \Bigr) \psi_{\sigma'\tau'}(\boldsymbol{k}) \Bigr\} . \nonumber \eqa
%
Here, $\psi_{\sigma\tau}(\boldsymbol{k})$ represents a four-component Dirac spinor, where $\sigma$ and $\tau$ are spin and chiral indexes, respectively.
$\bm{\sigma}_{\sigma\sigma'}$ and $\bm{\tau}_{\tau\tau'}$ are Pauli matrices acting on spin and ``orbital" spaces. The relativistic dispersion is represented in the chiral basis, where each eigen value of $\boldsymbol{\tau}_{\tau\tau'}^{z}$ expresses either $+$ or $-$ chirality, respectively. The mass term can be formulated as $m(|\boldsymbol{k}|) = m - \rho |\bm{k}|^{2}$, where $\mbox{sgn}(m) \mbox{sgn}(\rho) > 0$ corresponds to a topological insulating state while $\mbox{sgn}(m) \mbox{sgn}(\rho) < 0$ corresponds to a normal band insulating phase. $\mu$ is the chemical potential, controlled by doping. One may regard that this simplified effective model can be derived from a realistic band structure in Bi$_{1-x}$Sb$_{x}$, describing dynamics of electrons near the $\bm{L}$ point in momentum space.

It has been demonstrated that the mass gap can be tuned to vanish at $x = 3 \sim 4 ~ \%$ in Bi$_{1-x}$Sb$_{x}$, allowing us to reach the critical point between the topological and band insulating phases \cite{BiSb1,BiSb2,BiSb3}. It is straightforward to show that this gapless Dirac spectrum splits into a pair of Weyl points, breaking time reversal symmetry, for example, applying magnetic fields into the gapless semi-conductor
\bqa && H_{TRB} = g_{\psi} \psi_{\sigma\tau}^{\dagger}(\boldsymbol{k}) (\boldsymbol{B} \cdot \boldsymbol{\sigma}_{\sigma\sigma'} \otimes \boldsymbol{I}_{\tau\tau'}) \psi_{\sigma'\tau'}(\boldsymbol{k}) , \nonumber \eqa where $g_{\psi}$ is the Land\'{e} g-factor. The band touching point $(0,0,0)$ of the Dirac spectrum shifts into $(0,0, g_{\psi} B / v)$ and $(0,0,- g_{\psi} B / v)$ for each chirality along the direction of magnetic field, given by
\bqa && E_{\bm{k}} + \mu = \pm \sqrt{v^{2}[k_{x}^{2} + k_{y}^{2}] + [g_{\psi} B \pm v k_{z}]^{2}} . \nonumber \eqa Now, each spectrum is described by a two-component Weyl spinor with definite chirality, referred to as Weyl metal. See Fig. 1. One can also find this type of spectrum, breaking inversion symmetry instead of time reversal symmetry \cite{Weyl_Metal1,Weyl_Metal2,Weyl_Metal3}.

\begin{figure}[t]
\includegraphics[width=0.8\textwidth]{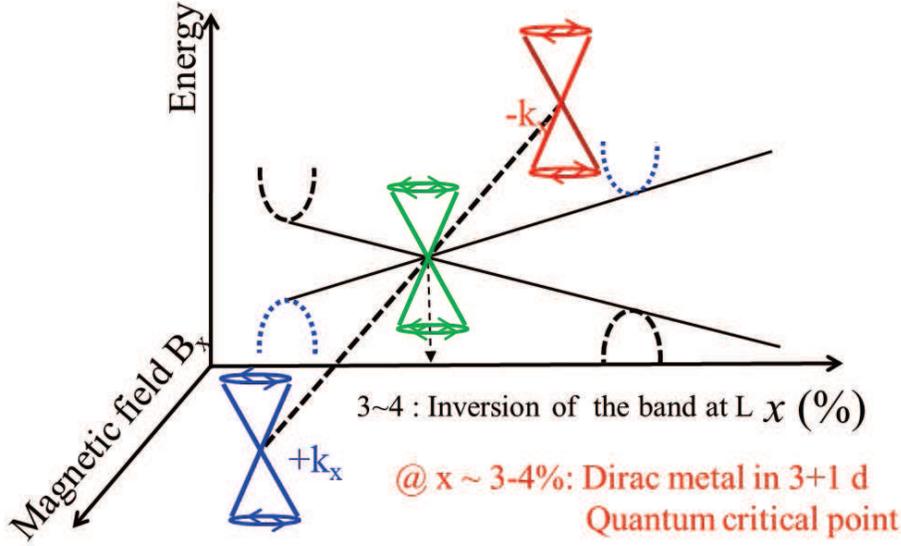}
\caption{A schematic phase diagram of Bi$_{1-x}$Sb$_{x}$ with a schematic band structure near the $\bm{L}$ point. A band inversion occurs around $x = 3 \sim 4 ~ \%$, giving rise to a gapless Dirac spectrum near the $\bm{L}$ point. Applying magnetic fields to this Dirac metal phase, the Dirac band splits into a pair of Weyl bands with definite chirality along the direction of the magnetic field, analogous to the polarization of a monopole and antimonopole pair induced by the magnetic field. (Figure from Kim et al. \cite{NLMR_Exp})} \label{Band_Structure}
\end{figure}

An important feature of Weyl metal results from the existence of Berry curvature on each Fermi surface \cite{Haldane}, referred to as a chiral Fermi surface since each Fermi surface encloses a Weyl point with definite chirality. When an electron moves on a chiral Fermi surface adiabatically, it feels an internal force given by Berry curvature \cite{Semiclassical_Eqs1,Semiclassical_Eqs2}. Considering that the Weyl Hamiltonian is given by $H_{W} = \bm{\sigma} \cdot \bm{p}$ with momentum $\bm{p}$ and spin $\bm{\sigma}$ near one Weyl point with $+$ chirality, the existence of the Berry curvature is quite analogous to dynamics of a spin under an adiabatic change of an external magnetic field $\bm{B}$ described by $H = \bm{\sigma} \cdot \bm{B}$ \cite{Sakurai_Textbook}. Recall that each Weyl point can be identified with a magnetic monopole in momentum space, regarded to be the source of Berry curvature \cite{Semiclassical_Eqs1,Semiclassical_Eqs2}. When both $+$ and $-$ magnetic charges exist at the same momentum point, there will be no net magnetic charge and no Berry curvature. Applying magnetic fields, each $\pm$ magnetic charge becomes ``polarized" in momentum space. However, unexpected and more interesting physics result beyond the presence of Berry curvature in the single chiral Fermi surface. First of all, dynamics of electron quasiparticles in one chiral Fermi surface is not free from that in the other paired chiral Fermi surface \cite{Anomaly_CFS}. This anomalous relation between the pair of Weyl points can be understood easily, considering one dimensional problems. At first, dynamics of electrons near a $+ k_{F}$ point looks independent against that near a $- k_{F}$ point, where $k_{F}$ is a Fermi momentum. However, they cannot be independent because there is a band structure to connect their dynamics. As a result, decrease in the number of $-$ chiral electrons must give rise to increase in that of $+$ chiral electrons, and vice versa \cite{Solid_State_Textbook}. Generally speaking, although symmetries or associated currents are conserved in the classical level, they are not respected in the quantum level sometimes, referred to as anomaly \cite{Anomaly_Textboook}. Here, chiral currents are not conserved quantum mechanically, thus the anomaly is called chiral anomaly \cite{Chiral_Anomaly}. This chiral anomaly also arises in three space dimensions when the band structure shows a pair of chiral Fermi surfaces (Weyl points), i.e., in Weyl metal \cite{Anomaly_CFS}.

A cautious person may point out that the band structure of Weyl metal is essentially the same as that of graphene, where the $+$ chirality Weyl spectrum at the $\bm{K}$ point and the $-$ chirality Weyl spectrum at the $-\bm{K}$ point allow us to call graphene two-dimensional Weyl metal \cite{Review_Graphene}. However, there is one critical difference between Weyl metal and graphene. Weyl electrons in the pair of Weyl points are not independent in Weyl metal as discussed above while they have ``nothing'' to do with each other in graphene. It is true that Weyl points in graphene can be regarded as a pair of Weyl points with opposite chirality according to the no-go theorem by Nielsen and Ninomiya \cite{NoGoTheorem1,NoGoTheorem2}. In addition, they can be shifted and merged into one Dirac point, applying effective ``magnetic'' fields to couple with the pseudo-spin of graphene \cite{Weyl_Metal1}. However, there does not exist such an anomaly relation between the pair of Weyl points in graphene, which means that currents are conserved separately for each Weyl cone in contrast with the case of Weyl metal as long as inter-valley scattering can be neglected. A crucial different aspect between two and three dimensions is that the irreducible representation of the Lorentz group is a four-component Dirac spinor in three dimensions while it is a two-component Weyl spinor in two dimensions \cite{Anomaly_Textboook}. As a result, the pair of Weyl points originates from the Dirac point in three dimensions, where such a pair of Weyl points is ``connected'' through the Dirac sea. On the other hand, each Weyl point of the pair exists ``independently'' in two dimensions. Chiral anomaly is the key feature of Weyl metal.

The above discussion defines our problem clearly. How should we modify the Drude model of the Landau's Fermi-liquid theory to incorporate the topological information of both the Berry curvature and chiral anomaly? How can we generalize the Landau's Fermi-liquid theory to encode the topological structure? In section II we discuss how the Drude model becomes generalized for a pair of chiral Fermi surfaces, based on Refs. \cite{NLMR_Exp} and \cite{NLMR}. This section covers an introduction to topological properties of the Weyl metallic state within the Boltzmann transport theory. In this respect we suggest readers, who are not much familiar to chiral anomaly, to check out this part carefully, combined with introduction. In section III we interpret recent experiments on Bi$_{1-x}$Sb$_{x}$ \cite{NLMR_Exp}, based on the Boltzmann transport theory with the modified Drude model, discussed in section II. Readers familiar to Weyl metal may focus on this section, skipping section II. 
%
%Since the comparison with experiments has been made briefly, we would like to refer more experimental details to the original article \cite{NLMR_Exp}. 
%
In section IV we discuss a topological Fermi-liquid theory, reviewing the Weyl Lagrangian in the first quantization which gives rise to the generalization of the Drude model. In particular, we justify our effective field theory for a pair of chiral Fermi surfaces, reproducing the enhancement of the longitudinal magneto-electrical conductivity given by the Boltzmann transport theory. Since this section is devoted to an aspect of quantum field theory, readers not much interested in the theoretical aspect may skip this section and proceed to the last section. Considering that this review article focuses on the introduction to topological properties of the Weyl metallic state in a pedagogical way as possible as it can, we suggest to see a recent preprint on the topological Fermi-liquid theory \cite{EFT} for rigorous derivations from a microscopic theory. In the last section we discuss a possible direction of this research, pointing out interesting problems on Weyl metal, where sections 5.2 and 5.3 are based on Refs. \cite{EFT} and \cite{WF_Law_WM}, respectively. In appendix we discuss the role of the topological-in-origin $\bm{E}\cdot\bm{B}$ term in the Weyl metallic phase. In particular, we prove that the angle coefficient $\theta(\bm{r})$ is given by $g_{\psi} \bm{B} \cdot \bm{r}$, where $g_{\psi}$ is the Land\'{e} g-factor and $\bm{B}$ is an external magnetic field. This tells us that electromagnetic properties of the Weyl metallic state are described not by conventional Maxwell equations but by axion electrodynamics \cite{Axion_EM}.

\section{How to describe a Weyl metallic state I: Boltzmann transport theory}

\subsection{The Boltzmann transport theory with the Drude model in the Landau's Fermi-liquid state}

The Boltzmann transport theory combined with the Drude model is our basic framework for metals described by Landau's Fermi-liquid theory \cite{LFL_Boltzmann}. The Boltzmann equation is given by \bqa && \Bigl( \frac{\partial}{\partial t} + \dot{\bm r} \cdot \bm{\nabla}_{\bm{r}} + \dot{\bm p} \cdot \bm{\nabla}_{\bm{p}} \Bigr) f(\bm{p};\bm{r},t) = I_{coll}[f(\bm{p};\bm{r},t)] , \nonumber \eqa where $f(\bm{p};\bm{r},t)$ is the distribution function with the conjugate momentum $\bm{p}$ of the relative coordinate and the center of mass coordinate ($\bm{r}$, $t$) in the Wigner transformation of the lesser Green's function \cite{Mahan_Book}. Based on the semi-classical wave-packet picture, one may regard that $\bm{p}$ is the momentum to characterize an internal state and $\dot{\bm r}$ is the group velocity of the particle \cite{Semiclassical_Eqs1,Semiclassical_Eqs2}. Dynamics of the internal momentum and the center of mass coordinate is described by the Drude model, \bqa && \boldsymbol{\dot{r}} = \frac{\partial \epsilon_{\boldsymbol{p}}}{\partial \boldsymbol{p}} , \nn && \boldsymbol{\dot{p}} = e \boldsymbol{E} + \frac{e}{c} \boldsymbol{\dot{r}} \times \boldsymbol{B} , \eqa which correspond to the group velocity with the band dispersion $\epsilon_{\boldsymbol{p}}$ and the equation of motion with the Lorentz force, respectively \cite{Solid_State_Textbook}. The right-hand side in the Boltzmann equation contains various collision terms, including electron correlations and impurity scattering effects. In this review article, we do not take into account electron correlations.

Incorporating the Drude model in the Boltzmann equation with impurity scattering, we obtain \bqa && \Bigl\{ \frac{\partial}{\partial t} + \bm{v}_{\bm{p}} \cdot \bm{\nabla}_{\bm{r}} + \Bigl( e \boldsymbol{E} + \frac{e}{c} \bm{v}_{\bm{p}} \times \boldsymbol{B} \Bigr) \cdot \bm{\nabla}_{\bm{p}} \Bigr\} f(\bm{p};\bm{r},t) \nn && = - \Gamma_{imp} [f(\bm{p};\bm{r},t) - f_{eq}(\bm{p})] - \int_{-\infty}^{t} d t' \alpha(t-t') [f(-\bm{p};\bm{r},t') - f_{eq}(\bm{p})] , \eqa where $\bm{v}_{\bm{p}}$ is the group velocity and $f_{eq}(\bm{p})$ is the equilibrium distribution function. The collision part consists of two types of impurity scattering contributions \cite{WAL_Boltzmann}. The first is an elastic impurity-scattering term in the relaxation-time approximation, where $\Gamma_{imp}^{-1} = (2 \pi n_{I} |V_{imp}|^{2} N_{F})^{-1}$ with an impurity concentration $n_{I}$, its potential strength $V_{imp}$, and the density of states $N_{F}$ corresponds to the mean free time, the time scale between events of impurity scattering. The second is a weak localization (weak antilocalization) term, expressed in a non-local way, which originates from multiple impurity scattering. $\alpha(t-t') = \pm \frac{\Gamma_{imp}}{\pi N_{F}} \int \frac{d^{3} \bm{q}}{(2\pi)^{3}} \exp\Bigl\{ - (D \bm{q}^{2} + \tau_{\phi}^{-1}) (t-t') \Bigr\}$ may be regarded as the diffusion kernel, which becomes more familiar, performing Fourier transformation as follows \bqa && \alpha(\nu) = \pm \int_{-\infty}^{t} d t' e^{i \nu(t-t')} \alpha(t-t') = \pm \frac{\Gamma_{imp}}{\pi N_{F}} \int \frac{d^{3} \bm{q}}{(2\pi)^{3}} \frac{1}{D \bm{q}^{2} - i \nu + \tau_{\phi}^{-1}} , \eqa where the sign of $+$ ($-$) represents the weak localization (weak antilocalization). $D$ is the diffusion coefficient and $N_{F}$ is the density of states at the Fermi energy. This expression is supplemented by the upper cut-off in the momentum integral, given by the reciprocal of the mean free path $\Gamma_{imp}/v_{F}$ with the Fermi velocity $v_{F}$, and $\tau_{\phi}$ corresponds to the lower cut-off, identified with the phase-coherence lifetime.

It is straightforward to solve this equation in the linear-response regime. Then, the electrical conductivity can be found from the following expression \bqa && \bm{j}(\bm{q},\nu) = \int \frac{d^{3} \bm{p}}{(2\pi)^{3}} \bm{v}_{\bm{p}} f(\bm{p};\bm{q},\nu) , \eqa where $\bm{q}$ and $\nu$ are momentum and frequency for the center of mass coordinate.

\subsection{A Boltzmann transport theory with a ``topological" Drude model}

The Weyl metallic state consists of a pair of chiral Fermi surfaces. As a result, the ``minimal'' model for Boltzmann equations should be constructed for each Fermi surface, given by \bqa && \Bigl( \frac{\partial}{\partial t} + \dot{\bm r}^{\chi} \cdot \bm{\nabla}_{\bm{r}} + \dot{\bm p}^{\chi} \cdot \bm{\nabla}_{\bm{p}} \Bigr) f^{\chi}(\bm{p};\bm{r},t) = I_{coll}[f^{\chi}(\bm{p};\bm{r},t)] , \nonumber \eqa where the superscript of $\chi$ represents each Weyl point with either $+$ or $-$ chirality.

An essential idea is to introduce the topological structure of both the Berry curvature and chiral anomaly into the Boltzmann equation framework \cite{Son_Boltzmann}, modifying the Drude model in an appropriate way for Weyl metal \cite{Semiclassical_Eqs1,Semiclassical_Eqs2}. We call it ``topological" Drude model, given by
\bqa && \boldsymbol{\dot{r}}^{\chi} = \frac{\partial \epsilon_{\boldsymbol{p}}^{\chi}}{\partial \boldsymbol{p}} + \boldsymbol{\dot{p}}^{\chi} \times \boldsymbol{b}_{\boldsymbol{p}}^{\chi} , \nn && \boldsymbol{\dot{p}}^{\chi} = e \boldsymbol{E} + \frac{e}{c} \boldsymbol{\dot{r}}^{\chi} \times \boldsymbol{B} , \eqa where $\boldsymbol{b}_{\boldsymbol{p}}^{\chi} = \bm{\nabla}_{\bm{p}} \times \bm{a}_{\bm{p}}^{\chi}$ is the Berry curvature and $\bm{a}_{\bm{p}}^{\chi} = i \langle u_{\bm{p}}^{\chi} | \bm{\nabla}_{\bm{p}} u_{\bm{p}}^{\chi} \rangle$ is the Berry connection with the Bloch's eigen state $| u_{\bm{p}}^{\chi} \rangle$. The emergence of the anomalous velocity term is a key feature in the topological Drude model, originating from the Berry curvature on the chiral Fermi surface. The Berry magnetic field is determined by \bqa && \bm{\nabla}_{\bm{p}} \cdot \bm{b}_{\bm{p}}^{\chi} = \chi \delta^{(3)}(\bm{p} - \bm{p}^{\chi}), \eqa where the ``magnetic" charge $\chi = \pm$ of a monopole in momentum space is given by the chirality of the Weyl point appearing at $\bm{p}^{\chi} = \chi g_{\psi} \bm{B}$ as discussed in the introduction.

It is straightforward to find the solution of these semi-classical equations of motion, given by \cite{Son_Boltzmann}
\bqa && \boldsymbol{\dot{r}}^{\chi} = \Bigl( 1 + \frac{e}{c} \boldsymbol{B} \cdot \boldsymbol{b}_{\boldsymbol{p}}^{\chi} \Bigr)^{-1} \Bigl\{ \boldsymbol{v}_{\boldsymbol{p}}^{\chi} + e \boldsymbol{E} \times \boldsymbol{b}_{\boldsymbol{p}}^{\chi} + \frac{e}{c} \boldsymbol{b}_{\boldsymbol{p}}^{\chi} \cdot \boldsymbol{v}_{\boldsymbol{p}}^{\chi} \boldsymbol{B} \Bigr\} , \nn && \boldsymbol{\dot{p}}^{\chi} = \Bigl( 1 + \frac{e}{c} \boldsymbol{B} \cdot \boldsymbol{b}_{\boldsymbol{p}}^{\chi} \Bigr)^{-1} \Bigl\{ e \boldsymbol{E} + \frac{e}{c} \boldsymbol{v}_{\boldsymbol{p}}^{\chi} \times \boldsymbol{B} + \frac{e^{2}}{c} (\boldsymbol{E} \cdot \boldsymbol{B}) \boldsymbol{b}_{\boldsymbol{p}}^{\chi} \Bigr\} . \eqa Here, $\boldsymbol{v}_{\boldsymbol{p}}^{\chi} = \bm{\nabla}_{\bm{p}} \epsilon_{\boldsymbol{p}}^{\chi}$ is the group velocity with a band structure $\epsilon_{\boldsymbol{p}}^{\chi}$. We would like to point out that this band structure needs not be linear-in-momentum strictly. We repeat that it is essential for a Fermi surface to enclose a Weyl cone. In the $\bm{\dot{r}}$ equation, both the second and third terms are anomalous velocity terms. In particular, the second term results in the anomalous Hall effect given by the Berry curvature \cite{AHE_Kubo1,AHE_Kubo2} and the third term gives rise to the chiral magnetic effect \cite{CME,CME1,CME2,CME3,CME_Kubo1,CME_Kubo2}. The last term in the $\bm{\dot{p}}$ equation is the source of chiral anomaly, responsible for the negative magneto-resistivity \cite{Anomaly_CFS,NLMR_Exp,NLMR,Son_Boltzmann}. In this paper we focus on the electrical-magneto resistivity.

Introducing the topological Drude model into the Boltzmann-equation framework, we reach the following expression
\bqa && \Bigl[ \frac{\partial}{\partial t} + \Bigl( 1 + \frac{e}{c} \boldsymbol{B} \cdot \boldsymbol{b}_{\boldsymbol{p}}^{\chi} \Bigr)^{-1} \Bigl\{ \boldsymbol{v}_{\boldsymbol{p}}^{\chi} + e \boldsymbol{E} \times \boldsymbol{b}_{\boldsymbol{p}}^{\chi} + \frac{e}{c} \boldsymbol{b}_{\boldsymbol{p}}^{\chi} \cdot \boldsymbol{v}_{\boldsymbol{p}}^{\chi} \boldsymbol{B} \Bigr\} \cdot \bm{\nabla}_{\bm{r}} \nn && + \Bigl( 1 + \frac{e}{c} \boldsymbol{B} \cdot \boldsymbol{b}_{\boldsymbol{p}}^{\chi} \Bigr)^{-1} \Bigl\{ e \boldsymbol{E} + \frac{e}{c} \boldsymbol{v}_{\boldsymbol{p}}^{\chi} \times \boldsymbol{B} + \frac{e^{2}}{c} (\boldsymbol{E} \cdot \boldsymbol{B}) \boldsymbol{b}_{\boldsymbol{p}}^{\chi} \Bigr\} \cdot \bm{\nabla}_{\bm{p}} \Bigr] f^{\chi}(\bm{p};\bm{r},t) \nn && = - \Gamma_{imp} [f^{\chi}(\bm{p};\bm{r},t) - f_{eq}(\bm{p})] - \Gamma_{imp}' [f^{\chi}(\bm{p};\bm{r},t) - f^{-\chi}(\bm{p};\bm{r},t)] \nn && - \int_{-\infty}^{t} d t' \alpha_{\chi}(t-t') [f^{\chi}(-\bm{p};\bm{r},t') - f_{eq}(\bm{p})] , \eqa where we introduced an inter Weyl-point scattering term into the Boltzmann equation phenomenologically with the relaxation rate $\Gamma_{imp}'$ for the inter-node scattering. The weak antilocalization kernel is given by \bqa && \alpha_{\chi}(\nu) = - \frac{\Gamma_{imp}+\Gamma'_{imp}}{\pi N_{F}} \int \frac{d^{3} \bm{q}}{(2\pi)^{3}} \frac{1}{D_{\chi} \bm{q}^{2} - i \nu + \tau_{\phi}^{-1}} , \nonumber \eqa where $D_{\chi}$ is the diffusion coefficient for each Weyl point, assumed to be identical, i.e., $D_{+} = D_{-} = D$.

It is also not much difficult to solve these coupled Boltzmann equations with the generalized Drude model. Then, electrical transport coefficients can be found from
\bqa && \bm{j}(\bm{q},\nu) = e \int \frac{d^{3} \bm{p}}{(2\pi)^{3}} \sum_{\chi = \pm} \Bigl( 1 + \frac{e}{c} \boldsymbol{B} \cdot \boldsymbol{b}_{\boldsymbol{p}}^{\chi} \Bigr) \dot{\bm r}_{\chi} f_{\chi}(\bm{p};\bm{q},\nu) \nn && = e \int \frac{d^{3} \bm{p}}{(2\pi)^{3}} \sum_{\chi = \pm} \Bigl\{ \boldsymbol{v}_{\boldsymbol{p}}^{\chi} + e \boldsymbol{E} \times \boldsymbol{b}_{\boldsymbol{p}}^{\chi} + \frac{e}{c} \boldsymbol{b}_{\boldsymbol{p}}^{\chi} \cdot \boldsymbol{v}_{\boldsymbol{p}}^{\chi} \boldsymbol{B} \Bigr\} f_{\chi}(\bm{p};\bm{q},\nu) . \eqa We would like to emphasize that the topological structure of both the Berry curvature and chiral anomaly modifies the definition of the electric current quite seriously, which turns out to be essential for anomalous transport phenomena in Weyl metal.

\subsection{Chiral anomaly in Weyl metal}

It is the next step to derive the hydrodynamic equation from the Boltzmann equation, performing the coarse graining procedure in momentum space. In normal metals described by the Boltzmann equation [Eq. (2)] with the conventional Drude model [Eq. (1)], we find the current conservation law, given by \bqa && \frac{\partial N}{\partial t} + \bm{\nabla}_{\bm{r}} \cdot \bm{j} = 0 , \eqa where $N = \int_{-\infty}^{\infty} d \epsilon \rho(\epsilon) f(\epsilon;\bm{r},t)$ is an electron density with the density of states $\rho(\epsilon) = \int \frac{d^{3} \bm{p}}{(2\pi)^{3}} \delta(\epsilon_{\bm{p}} - \epsilon)$ and $\bm{j} = \int \frac{d^{3} \bm{p}}{(2\pi)^{3}} \bm{v}_{\bm{p}} f(\bm{p};\bm{r},t)$ is an electric current, contributed dominantly from electrons near the Fermi surface. We recall that collision terms cannot play the role of either sources or sinks for currents \cite{LFL_Boltzmann}.

However, the current conservation law breaks down around each chiral Fermi surface \cite{Anomaly_CFS}, regarded to be gauge anomaly \cite{Anomaly_Textboook}. Multiplying $\Bigl( 1 + \frac{e}{c} \boldsymbol{B} \cdot \boldsymbol{b}^{\chi}_{\boldsymbol{p}} \Bigr)$ to the Boltzmann equation [Eq. (8)] with the topological Drude model [Eq. (7)], we obtain \cite{Son_Boltzmann}
\bqa && \frac{\partial N^{\chi}}{\partial t} + \bm{\nabla}_{\bm{r}} \cdot \bm{j}^{\chi} = k^{\chi} \frac{e^{2}}{4\pi^{2}} \bm{E} \cdot \bm{B} , \eqa where $N^{\chi} = \int_{-\infty}^{\infty} d \epsilon \rho^{\chi}(\epsilon) f^{\chi}(\epsilon;\bm{r},t)$ is an electron density with $\rho^{\chi}(\epsilon) = \int \frac{d^{3} \bm{p}}{(2\pi)^{3}} \Bigl( 1 + \frac{e}{c} \boldsymbol{B} \cdot \boldsymbol{b}^{\chi}_{\boldsymbol{p}} \Bigr) \delta(\epsilon_{\bm{p}} - \epsilon)$ and $\bm{j}^{\chi} = \int \frac{d^{3} \bm{p}}{(2\pi)^{3}} \Bigl( 1 + \frac{e}{c} \boldsymbol{B} \cdot \boldsymbol{b}^{\chi}_{\boldsymbol{p}} \Bigr) \bm{\dot{r}}^{\chi} f^{\chi}(\bm{p};\bm{r},t) = \int \frac{d^{3} \bm{p}}{(2\pi)^{3}} \Bigl\{ \boldsymbol{v}_{\boldsymbol{p}} + e \boldsymbol{E} \times \boldsymbol{b}^{\chi}_{\boldsymbol{p}} + \frac{e}{c} \boldsymbol{b}^{\chi}_{\boldsymbol{p}} \cdot \boldsymbol{v}_{\boldsymbol{p}} \boldsymbol{B} \Bigr\} f^{\chi}(\bm{p};\bm{r},t)$ is an electric current around each chiral Fermi surface. $k^{\chi} = \frac{1}{2\pi} \int d \bm{S}_{\bm{p}} \cdot \boldsymbol{b}^{\chi}_{\boldsymbol{p}} = \chi$ is a magnetic charge at each Weyl point. It is clear that the breakdown of the current conservation law around each Weyl point results from the $\bm{E}\cdot\bm{B}$ term, introduced by the modification of the Drude model. This gauge anomaly around each Weyl point becomes canceled by the existence of its partner. The $+$ chiral charge plays the role of a source in this hydrodynamic equation while the $-$ chiral charge does that of a sink. As a result, the total current is conserved, given by
\bqa && \frac{\partial (N^{+} + N^{-})}{\partial t} + \bm{\nabla}_{\bm{r}} \cdot (\bm{j}^{+}+\bm{j}^{-}) = 0  \eqa while the chiral current is not, described by
\bqa && \frac{\partial (N^{+} - N^{-})}{\partial t} + \bm{\nabla}_{\bm{r}} \cdot (\bm{j}^{+}-\bm{j}^{-}) = \frac{e^{2}}{2\pi^{2}} \bm{E} \cdot \bm{B} . \eqa This is the chiral anomaly in three dimensions.

\section{Anomalous transport phenomena in Weyl metal}

An effective Boltzmann transport theory has been constructed for a pair of chiral Fermi surfaces, generalizing the Drude model to incorporate both the Berry curvature and chiral anomaly. It is straightforward to solve the pair of Boltzmann equations. Here, we do not present details of calculations in the Boltzmann equation approach. Instead, we focus on novel physics, arising from the chiral anomaly. We refer all details of the Boltzmann equation approach to Ref. \cite{NLMR}.

\subsection{$\bm{E} = E \hat{\bm x}$ and $\bm{B} = B \hat{\bm z}$}

In the transverse setup of $\bm{E} = E \hat{\bm x}$ with $\bm{B} = B \hat{\bm z}$, the electrical-magneto conductivity is given by
\bqa && \sigma_{xx} \approx 2 \sigma \frac{1 + \alpha/\Gamma_{imp} + \Gamma_{imp}' / \Gamma_{imp}}{ [1 + \alpha/\Gamma_{imp}]^{2} + \omega_{c}^{2}/\Gamma_{imp}^{2}} . \eqa Here, $\sigma \approx \frac{e^{2}}{\Gamma_{imp}} \int \frac{d^{3} \bm{p}}{(2\pi)^{3}} \Bigl( - \frac{\partial }{\partial \epsilon} f_{eq} (\epsilon) \Bigr) \frac{|\bm{v}_{\bm{p}}|^{2}}{3}$ is the Drude conductivity for each Weyl cone, and $\omega_{c}$ is the cyclotron frequency of the Weyl electron on the chiral Fermi surface. $2$ comes from the pair of Weyl cones. $\alpha = - \frac{\Gamma_{imp}}{\pi N_{F}} \int \frac{d^{3} \bm{q}}{(2\pi)^{3}} \frac{1}{D \bm{q}^{2} + \tau_{\phi}^{-1}}$ is the kernel of weak antilocalization in the dc limit. This expression reads \bqa && \rho_{xx} \approx \Bigl( 1 - \frac{\Gamma_{imp}'}{\Gamma_{imp}} \Bigr) \rho - \mathcal{C} e^{2} N_{F} \rho^{2} \int^{1/l_{imp}}_{1/l_{ph}} d q q^{2} \frac{1}{ \bm{q}^{2}} \eqa in the leading order for magnetic fields. $\rho = \sigma^{-1}$ is a residual resistivity due to elastic impurity scattering and $\mathcal{C}$ is a positive numerical constant. $l_{imp}$ in the upper cut-off is the mean-free path and $l_{ph}$ is the phase-coherence length. If one sets $l_{ph}^{-1} \propto \sqrt{B}$ in the lower cut-off, we reproduce the magneto-resistivity with weak antilocalization in three dimensions \cite{Review_Disorder}.

The Hall conductivity is \bqa && \sigma_{yx} = - e^{2} \int \frac{d^{3} \bm{p}}{(2\pi)^{3}} [b_{z}^{+}(\bm{p}) + b_{z}^{-}(\bm{p})] f_{eq}(\bm{p}) \nn && - 2 \sigma \frac{ \omega_{c} / \Gamma_{imp} }{ (1 + \alpha/\Gamma_{imp})^{2} + (\omega_{c}/\Gamma_{imp})^{2} } \frac{1 + \alpha/\Gamma_{imp} + \Gamma_{imp}' / \Gamma_{imp}}{1 + \alpha/\Gamma_{imp}} , \eqa where the first term is an anomalous contribution resulting from the Berry curvature and the second is the normal Hall contribution from the pair of Fermi surfaces with weak antilocalization. Inserting $\bm{b}_{\bm{p}}^{\chi} \propto \chi \frac{\bm{\hat{p}}}{|\bm{p} - \chi g_{\psi} \bm{B}|^{2}}$ with $\bm{B} = B \bm{\hat{z}}$ and $\chi = \pm$ into the expression of the anomalous Hall coefficient and performing the momentum integration, we find that it is proportional to the momentum-space distance between the pair of Weyl points \cite{NLMR}, i.e., $g_{\psi} B$, consistent with that based on the diagrammatic analysis \cite{AHE_Kubo1,AHE_Kubo2}. For the normal contribution, the presence of the inter-node scattering modifies the Hall coefficient as follows
\bqa && \rho_{yx} = \frac{\sigma_{yx}}{\sigma_{xx}^{2} + \sigma_{yx}^{2}} = - \frac{1}{n e c} \Bigl(1 + \frac{\Gamma_{imp}'}{\Gamma_{imp}} \frac{1}{1 + \alpha/\Gamma_{imp}} \Bigr) , \eqa which turns out to be not a constant but a function of magnetic field, combined with the weak antilocalization correction.

\subsection{$\bm{E} = E \hat{\bm x}$ and $\bm{B} = B \hat{\bm x}$}

In the longitudinal setup of $\bm{E} = E \hat{\bm x}$ with $\bm{B} = B \hat{\bm x}$, we obtain the longitudinal electrical-magneto conductivity given by
\bqa && \sigma_{xx} = 2 \sigma \Bigl\{ 1 + \mathcal{C}_{ABJ} \Bigl(\frac{e B}{c}\Bigr)^{2} + \frac{\Gamma_{imp}'}{\Gamma_{imp}} \frac{1}{1 + \alpha/\Gamma_{imp}} \Bigr\} \frac{1}{1 + \alpha/\Gamma_{imp}} \nn && - \frac{4}{3} \sigma \mathcal{C}_{ABJ} m^{2} \omega_{c}^{2} \Bigl( \frac{\omega_{c}^{2} / \Gamma_{imp}^{2}}{1 + \alpha / \Gamma_{imp}} + \frac{\Gamma_{imp}'}{\Gamma_{imp}} \Bigr) \frac{ 1 }{ [1 + \alpha / \Gamma_{imp}]^{2} + \omega_{c}^{2} / \Gamma_{imp}^{2} } \eqa with
\bqa && \mathcal{C}_{ABJ} \approx \frac{e^{2}}{\Gamma_{imp}} \int \frac{d^{3} \bm{p}}{(2\pi)^{3}} \Bigl( - \frac{\partial }{\partial \epsilon} f_{eq} (\epsilon) \Bigr) \frac{|\bm{v}_{\bm{p}}|^{2}}{3} |\bm{b}_{\bm{p}}|^{2} \Bigl/ \frac{e^{2}}{\Gamma_{imp}} \int \frac{d^{3} \bm{p}}{(2\pi)^{3}} \Bigl( - \frac{\partial }{\partial \epsilon} f_{eq} (\epsilon) \Bigr) \frac{|\bm{v}_{\bm{p}}|^{2}}{3} . \nonumber \eqa $\omega_{c}$ is the ``cyclotron" frequency with the magnetic field of the $x-$direction. If we take the limit of $\Gamma_{imp}' / \Gamma_{imp} \rightarrow 0$, this expression is further simplified as \bqa && \sigma_{xx} = 2 \sigma \Bigl\{ 1 + \mathcal{C}_{ABJ} \Bigl(\frac{e B}{c}\Bigr)^{2} \Bigr\} \frac{1}{1 + \alpha/\Gamma_{imp}} \nn && - \frac{4}{3} \sigma \mathcal{C}_{ABJ} m^{2} \omega_{c}^{2} \frac{1}{1 + \alpha / \Gamma_{imp}} \frac{ \omega_{c}^{2} / \Gamma_{imp}^{2} }{ [1 + \alpha / \Gamma_{imp}]^{2} + \omega_{c}^{2} / \Gamma_{imp}^{2} } . \nonumber \eqa Focusing on the low-field region, we obtain \bqa && \sigma_{xx} = 2 \sigma \Bigl\{ 1 + \mathcal{C}_{ABJ} \Bigl(\frac{e B}{c}\Bigr)^{2} \Bigr\} \frac{1}{1 + \alpha/\Gamma_{imp}} , \nonumber \eqa referred to as the enhancement of the longitudinal magneto-conductivity, where the $B^{2}$ contribution results from the $\bm{E}\cdot\bm{B}$ term \cite{NLMR_Exp,Anomaly_CFS,Son_Boltzmann,NLMR}. Inserting the weak antilocalization correction into the above formula and considering $l_{ph}^{-1} = (\mathcal{C}'/\mathcal{C}) \sqrt{B}$ with a positive constant $\mathcal{C}'$, we find \bqa && \sigma_{xx} = \frac{2}{\rho_{imp}} \Bigl\{ 1 + \mathcal{C}_{ABJ} \Bigl(\frac{e B}{c}\Bigr)^{2} \Bigr\} \frac{1}{1 - \mathcal{C} e^{2} N_{F} \rho_{imp} l_{imp}^{-1} + \mathcal{C}' e^{2} N_{F} \rho_{imp} \sqrt{B}} , \label{LMR_ABJ} \eqa which turns out to fit the experimental data well \cite{NLMR_Exp}.

It is not that difficult to understand intuitively why the chiral anomaly enhances the longitudinal conductivity. Chiral anomaly allows the dissipationless current channel between the pair of Weyl points, reducing electrical resistivity along the direction of the applied magnetic field \cite{Anomaly_CFS}. Such an effect appears to be the $B^{2}$ term in front of the ``normal" resistivity.

Following the same strategy as that of the magneto-conductivity, it is straightforward to find the Hall conductivity around each Weyl point, which consists of anomalous contributions. One is the anomalous Hall effect resulting from the Berry curvature like the first term of Eq. (16) but with $\bm{B} = B \bm{\hat{x}}$ while the other is another type of the anomalous Hall effect originating from the chiral anomaly, where the topological $\bm{E}\cdot\bm{B}$ term gives rise to an additional force beyond the conventional Lorentz force. However, we find that the anomaly-induced anomalous Hall effect does not exist, inserting $\bm{b}_{\bm{p}}^{\chi} \propto \chi \frac{\bm{\hat{p}}}{|\bm{p} - \chi g_{\psi} \bm{B}|^{2}}$ with $\bm{B} = B \bm{\hat{x}}$ into the expression and performing the momentum integration. The ``conventional" anomalous Hall effect also vanishes. In other words, the Hall coefficient turns out to vanish identically in this longitudinal setup \cite{NLMR}.

\subsection{Comparison with experiments}

\begin{figure}[t]
\includegraphics[width=0.5\textwidth]{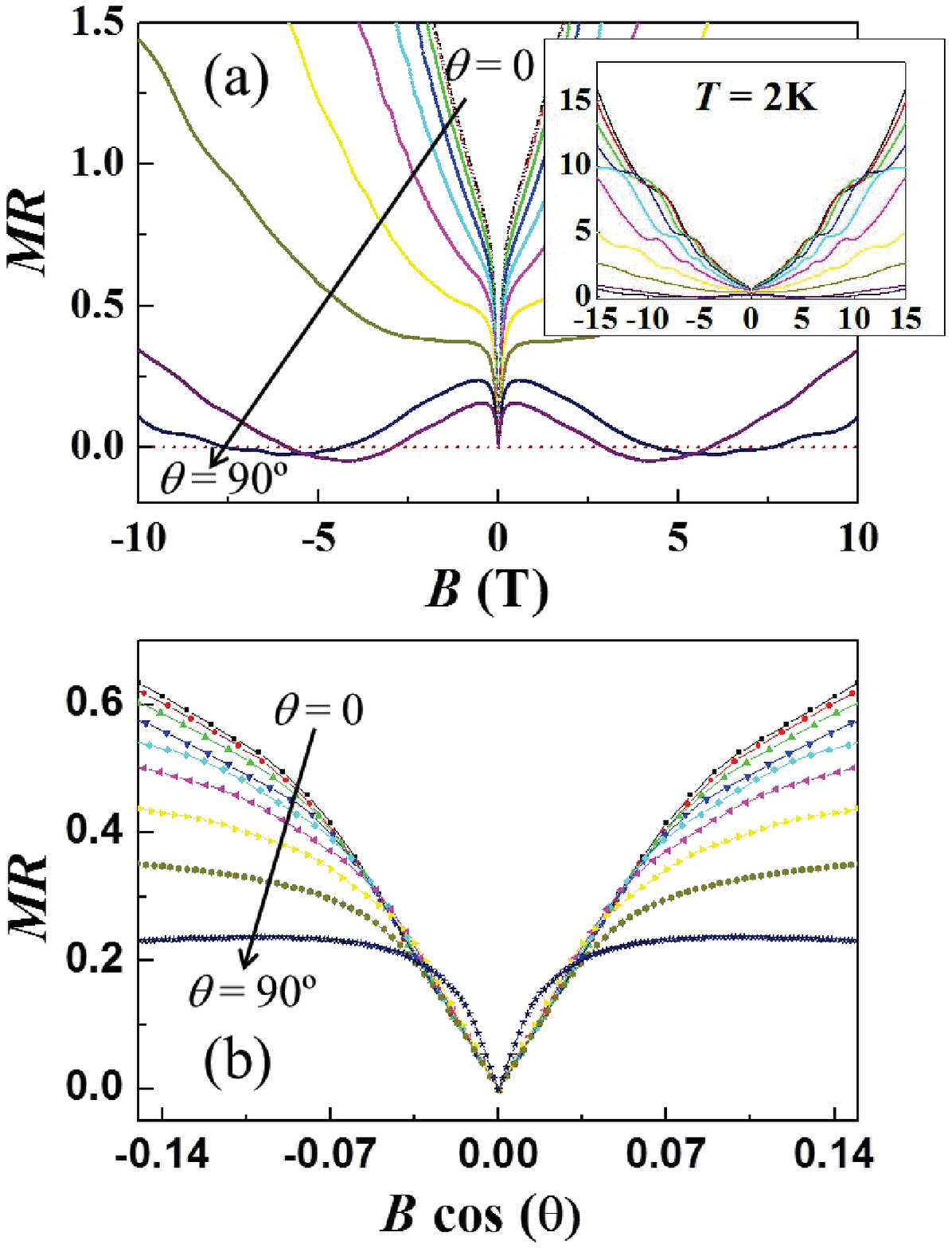}
\includegraphics[width=0.55\textwidth]{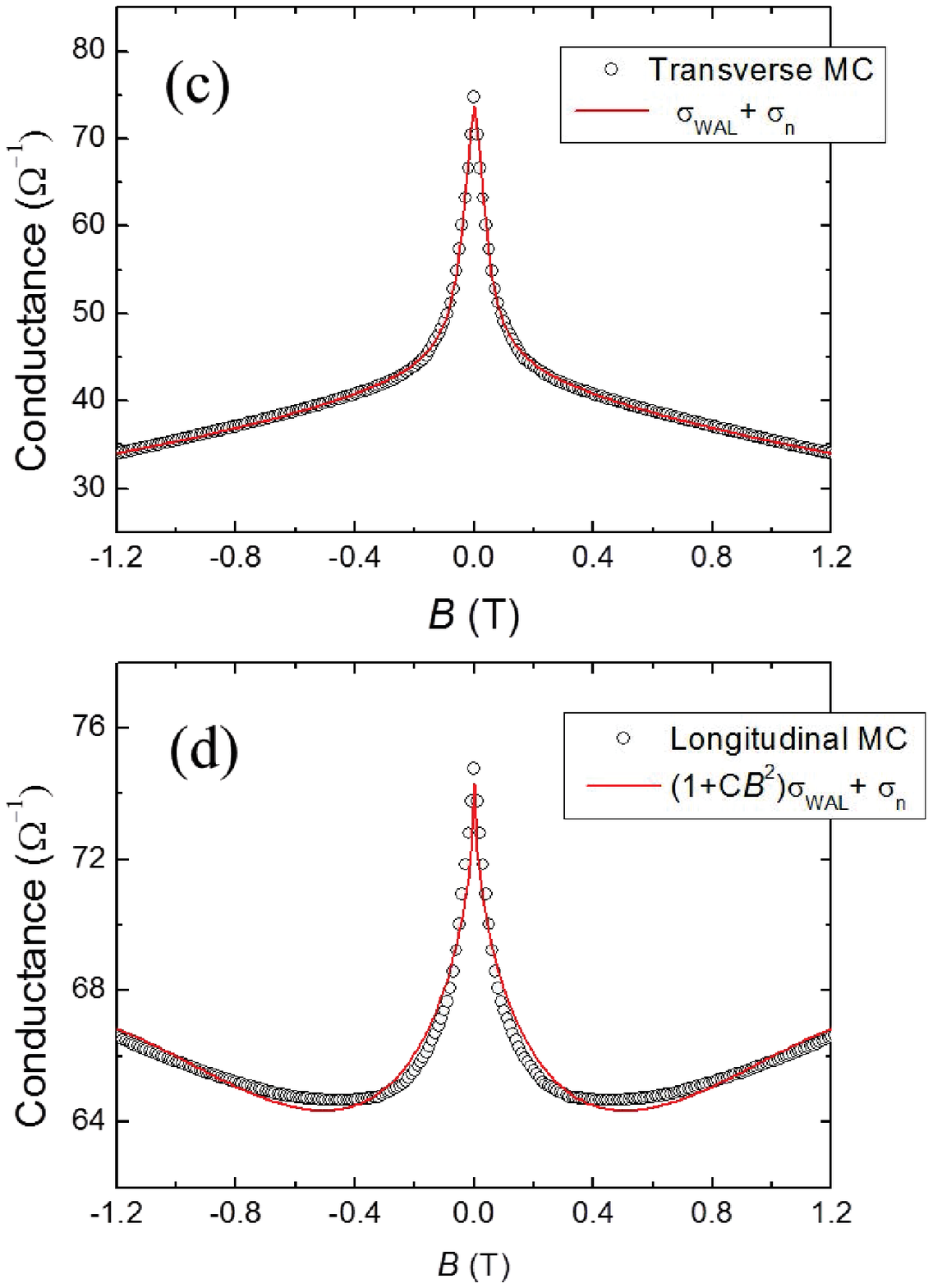}
\caption{(a) Angle dependence of electrical-magneto resistivity for Bi$_{0.97}$Sb$_{0.03}$ single crystals. $\theta = 0$ corresponds to the transverse setup of $\bm{E} = E \hat{\bm x}$ with $\bm{B} = B \hat{\bm z}$ while $\theta = 90$ corresponds to the longitudinal setup of $\bm{E} = E \hat{\bm x}$ with $\bm{B} = B \hat{\bm x}$. Electrical resistivity becomes suppressed dramatically in the longitudinal setup, resulting from the chiral anomaly between the pair of chiral Fermi surfaces. (b) Three dimensional weak antilocalization. Breakdown of scaling for the magnetic field of the $z-$direction implies that this weak antilocalization appears from not the surface state but the bulk state. Indeed, $\sqrt{B}$ fits this data quite well. (c) Fitting of the transverse electrical-magneto conductivity based on Eq. (19) without the $B^{2}$ term and (d) Fitting of the longitudinal electrical-magneto conductivity based on Eq. (19) with the $B^{2}$ term. We also introduced the normal contribution from electrons of the Fermi surface near the $\bm{T}$ point. For details, see the text. (Figure from Kim et al. \cite{NLMR_Exp})} \label{LMR_Angle_Dependence_Fitting}
\end{figure}

In order to explain the experimental data of Ref. \cite{NLMR_Exp}, we introduced two contributions for magneto-conductivity, where one results from Weyl electrons near the $\bm{L}-$point of the momentum space and the other comes from normal electrons near the $\bm{T}-$point. Subtracting out the cyclotron contribution of normal electrons in the transverse setup ($\bm{B} \perp \bm{E}$), we could fit the data based on the three-dimensional weak-antilocalization formula, given by Weyl electrons, where the weak-antilocalization correction has been Taylor-expanded for the weak-field region below $1.2$ T. On the other hand, the cyclotron contribution around the $\bm{T}-$point almost vanishes for the longitudinal setup ($\bm{B} \parallel \bm{E}$) as it must be, and the residual resistivity for normal electrons is almost identical with that of the transverse setup. Subtracting out the $\bm{T}-$point contribution, we could fit the data with Eq. (\ref{LMR_ABJ}) in the regime of the weak magnetic field below $1.2$ T, where the weak-antilocalization correction has been also Taylor-expanded. Again, the weak-antilocalization correction turns out to be almost identical with that of the transverse setup while we have an additional constant $\mathcal{C}_{ABJ}$ in the longitudinal setup, the origin of which is the chiral anomaly. See Fig. 2.

Recently, we investigated effects of thermal fluctuations on this chiral-anomaly driven enhancement of the longitudinal electrical-magneto conductivity, where temperature dependence of the $B^{2}-$enhancement factor $C_{w}$ has been measured. See Fig. 3. First of all, we found that the temperature dependence of the enhancement factor follows the standard mean-field behavior like an order parameter, i.e., $C_{w} \propto (1 - T/T_{c})^{1/2}$ with $T_{c} \sim 110$ K, quite unexpected. We believe that this measurement proposes an interesting question how thermal fluctuations can affect some topological properties, not much discussed until now.

\begin{figure}[here]
\includegraphics[width=0.9\textwidth]{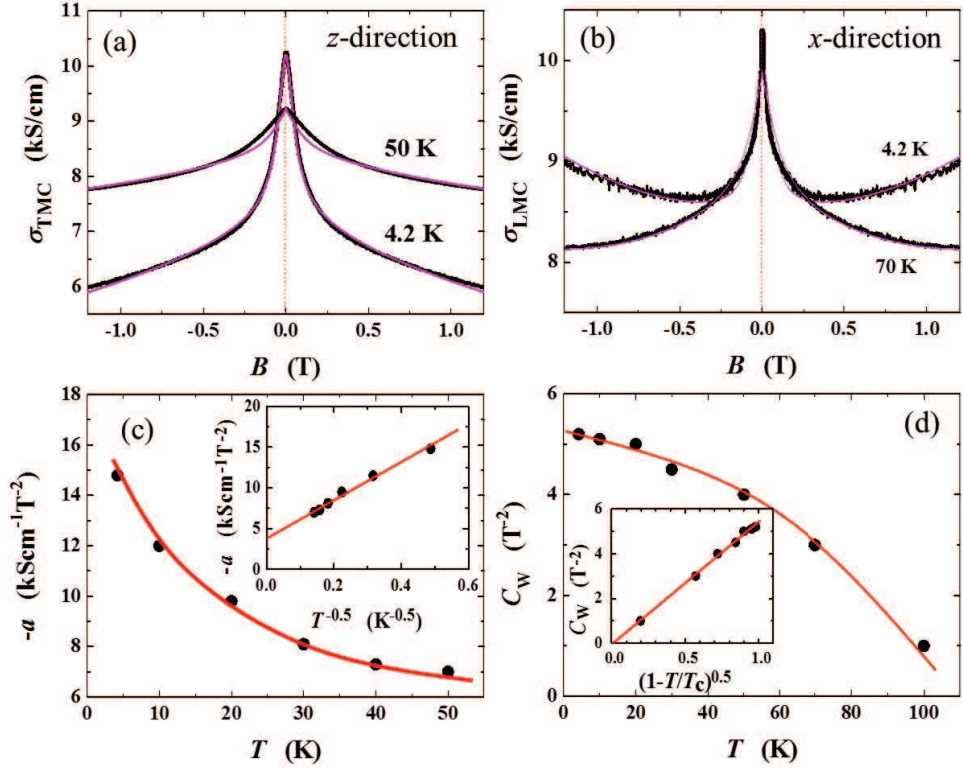}
\caption{(a) Temperature dependence of transverse electrical-magneto conductivity (TMC: $\bm{E} = E \hat{\bm x}$ and $\bm{B} = B \hat{\bm z}$) for Bi$_{0.97}$Sb$_{0.03}$ single crystals, well fitted by $\sigma_{TMC} = \sigma_{WAL} + \sigma_{n}$ with $\sigma_{WAL} = a \sqrt{B} + \sigma_{0}$ and $\sigma_{n} = (A B^{2} + \rho_{0})^{-1}$, as discussed in the text. The weak anti-localization (WAL) contribution observed in $\sigma_{TMC}$ is reduced gradually, increasing temperature, where its magnitude is reduced about $1/5$ at 50 K, compared to that at 4.2 K. (b) Temperature dependence of longitudinal electrical-magneto conductivity (LMC: $\bm{E} = E \hat{\bm x}$ and $\bm{B} = B \hat{\bm x}$) for Bi$_{0.97}$Sb$_{0.03}$ single crystals, well fitted by $\sigma_{LMC} = (1 + C_{w} B^{2}) \sigma_{WAL} + \sigma_{n}$. We would like to point out that the weak anti-localization contribution remains more robust than that of TMC, expected to result from the absence of an orbital effect in the longitudinal setup. (c) Temperature dependence of the weak anti-localization contribution in TMC. The coefficient $a$ is found proportional to $- T^{-1/2}$, which completes the scaling relation of $\sigma_{WAL} \propto - \sqrt{B/T}$ in the three-dimensional weak anti-localization correction. (d) Temperature dependence of the enhancement factor of the $B^{2}$ term $C_{w}$ in LMC. It is quite unexpected that the enhancement factor follows the standard mean-field behavior like an order parameter, i.e., $C_{w} \propto (1 - T/T_{c})^{1/2}$ with $T_{c} \sim 110$ K. These measurements have been performed in Wuhan National High Magnetic Field Center.} \label{LMR_Temperature_Dependence_Fitting}
\end{figure}

\section{How to describe a Weyl metallic state II: Topological Fermi-liquid theory}

\subsection{An effective field theory for a pair of chiral Fermi surfaces}

\subsubsection{First quantization: Derivation of the topological Drude model}

In order to construct a topological Fermi-liquid theory, we start from dynamics of one Weyl electron. The time evolution operator is given by
\bqa && \langle x_{f} |  e^{i \hat{H}_{W} (t_{f} - t_{i})} | x_{i} \rangle = \int_{x_{i}}^{x_{f}} D x D p \mathcal{P} \exp \Bigl\{ i \int_{t_{i}}^{t_{f}} d t (\bm{p} \cdot \dot{\bm{x}} - \bm{\sigma} \cdot \bm{p}) \Bigr\} \nonumber \eqa in the path integral representation. $(x_{i(f)}, t_{i(f)})$ denotes an initial (final) point in the configuration space, and $\mathcal{P}$ means ``path ordering". The Weyl Hamiltonian is given by $H_{W} = \bm{\sigma} \cdot \bm{p}$ for a Weyl point with $+$ chirality, where $\bm{\sigma}$ is a pauli matrix representing spin.

We focus on the case when the chemical potential lies away from the Weyl point, allowing a Fermi surface. We derive an effective Hamiltonian for dynamics of such a Weyl electron on the Fermi surface. An idea is to reformulate the Weyl Hamiltonian as follows, resorting to the CP$^{1}$ representation \cite{CME2} \bqa && H_{W} = \bm{\sigma} \cdot \bm{p} = |\bm{p}| \bm{V}_{\bm{p}} \bm{\sigma}_{3} \bm{V}_{\bm{p}}^{\dagger} , \nonumber \eqa where $\bm{V}_{\bm{p}} = \left( \begin{array} [c]{cc} z_{\bm{p}\uparrow} & z_{-\bm{p}\downarrow}^{\dagger} \\ z_{\bm{p}\downarrow} & - z_{-\bm{p}\uparrow}^{\dagger} \end{array} \right)$ is a two by two matrix with a unimodular constraint of $z_{\bm{p}\uparrow}^{\dagger} z_{\bm{p}\uparrow} + z_{\bm{p}\downarrow}^{\dagger} z_{\bm{p}\downarrow} = 1$. This is nothing but a change of basis which diagonalizes the Weyl Hamiltonian. The path integral representation becomes also reformulated as follows \cite{CME2} \bqa && \langle x_{f} |  e^{i \hat{H}_{W} (t_{f} - t_{i})} | x_{i} \rangle = \bm{V}_{p_{f}} \int_{x_{i}}^{x_{f}} D x D p \mathcal{P} \exp \Bigl\{ i \int_{t_{i}}^{t_{f}} d t (\bm{p} \cdot \dot{\bm{x}} - |\bm{p}| \bm{\sigma}_{3} - \hat{\bm{a}}_{\bm{p}} \cdot \dot{\bm{p}}) \Bigr\} \bm{V}_{\bm{p}_{i}} , \nonumber \eqa where $\bm{V}_{\bm{p}_{i(f)}}$ is an SU(2) matrix at an initial (final) position. As clearly seen, we have a Berry phase term within this diagonal basis \bqa && \mathcal{S}_{B} = - i \int_{t_{i}}^{t_{f}} d t \hat{\bm{a}}_{\bm{p}} \cdot \dot{\bm{p}} = - i \int_{\mathcal{C}} d \bm{p} \cdot \hat{\bm{a}}_{\bm{p}} , \eqa where $\mathcal{C}$ is a path of a Weyl electron adiabatically moving on the Fermi surface and \bqa && \hat{\bm{a}}_{\bm{p}} = i \bm{V}_{\bm{p}}^{\dagger} \bm{\nabla}_{\bm{p}} \bm{V}_{\bm{p}} \nonumber \eqa is an SU(2) Berry gauge field.

Introducing electromagnetic fields into the Weyl Lagrangian and reformulating it based on the diagonalized basis more carefully, one finds an effective action which describes dynamics of a Weyl electron on the Fermi surface \cite{QBE_WM3,CEFT_QED4_1,CEFT_QED4_2} \bqa && S_{eff} = \int_{t_{i}}^{t_{f}} d t \Bigl\{ \bm{p} \cdot \dot{\bm{x}} + \bm{A} \cdot \dot{\bm{x}} - \Phi - (|\bm{p}| - \bm{b}_{\bm{p}} \cdot \bm{B}) - \bm{a}_{\bm{p}} \cdot \dot{\bm{p}} \Bigr\} , \eqa where $\bm{A}$ and $\Phi$ are electromagnetic vector potential and scalar potential, respectively, and the $11$ component of the SU(2) Berry gauge field is chosen to be \bqa && \bm{a}_{\bm{p}} = i z_{\bm{p}\sigma}^{\dagger} \bm{\nabla}_{\bm{p}} z_{\bm{p}\sigma} \eqa and $\boldsymbol{b}_{\boldsymbol{p}} = \bm{\nabla}_{\bm{p}} \times \bm{a}_{\bm{p}}$ is the Berry curvature field near the Fermi surface.

It is straightforward to find a set of semi-classical equations of motion, given by \cite{CME1,CME2} \bqa && \dot{\bm{x}} = \bm{v}_{\bm{p}} + \dot{\bm{p}} \times \bm{b}_{\bm{p}} , \nn && \dot{\bm{p}} = \bm{E} + \dot{\bm{x}} \times \bm{B} , \eqa where $\bm{v}_{\bm{p}} = \bm{\nabla}_{\bm{p}} (|\bm{p}| - \bm{b}_{\bm{p}} \cdot \bm{B})$ is the group velocity, renormalized by the relativistic correction of the magnetic-moment coupling. This is the topological Drude model that we discussed in section II. The solution is \bqa && \sqrt{G_{\bm{p}}} \dot{\bm{x}} = \bm{v}_{\bm{p}} + \bm{E} \times \bm{b}_{\bm{p}} + \bm{B} (\bm{v}_{\bm{p}} \cdot \bm{b}_{\bm{p}}) , \nn && \sqrt{G_{\bm{p}}} \dot{\bm{p}} = \bm{E} + \bm{v}_{\bm{p}} \times \bm{B} + \bm{b}_{\bm{p}} (\bm{E} \cdot \bm{B}) , \eqa where $\sqrt{G_{\bm{p}}} = (1 + \bm{b}_{\bm{p}} \cdot \bm{B})$ is a measure of the phase space with chiral charge.

\subsubsection{Second quantization: Toward a topological Fermi-liquid theory}

The next step is to reformulate the first-quantized effective Weyl action in the second quantization, which describes dynamics of many electrons near the Fermi surface to enclose a Weyl point. An idea is to introduce a spinless fermion field at the chiral Fermi surface, related with an original Weyl electron via the CP$^{1}$ representation as follows \bqa && \left( \begin{array} [c]{c} c_{\uparrow}(\bm{p}) \\ c_{\downarrow}(\bm{p}) \end{array} \right) = \left( \begin{array} [c]{cc} z_{\bm{p}\uparrow} & z_{-\bm{p}\downarrow}^{\dagger} \\ z_{\bm{p}\downarrow} & - z_{-\bm{p}\uparrow}^{\dagger} \end{array} \right) \left( \begin{array} [c]{cc} \psi(\bm{p}) \\ \phi(\bm{p}) \end{array} \right) . \eqa Since we focus on low energy dynamics near the chiral Fermi surface, we find the relation between the spinless fermion field and original Weyl electron field, given by \bqa && c_{\sigma}(\bm{p}) = z_{\bm{p}\sigma} \psi(\bm{p}) . \eqa

Resorting to this slave-fermion representation for Weyl metal, we can reformulate the gauge-matter coupling term in the imaginary-time formulation at finite temperature as follows
\bqa && \mathcal{S}_{G} = \int_{0}^{\beta} d \tau \int \frac{d^{3} \bm{p}}{(2\pi)^{3}} \sqrt{G_{\bm{p}}} \int \frac{d^{3} \bm{q}}{(2\pi)^{3}} \bm{A}_{\bm{q}} \cdot c_{\sigma}^{\dagger}(\bm{p}+\bm{q}) \dot{\bm{x}}_{\bm{p},\bm{q}} c_{\sigma}(\bm{p}) \nn && \longrightarrow \int_{0}^{\beta} d \tau \int \frac{d^{3} \bm{p}}{(2\pi)^{3}} \sqrt{G_{\bm{p}}} \int \frac{d^{3} \bm{q}}{(2\pi)^{3}} [z_{\chi\bm{p}+\bm{q}\uparrow}^{\dagger} z_{\chi\bm{p}\uparrow} + z_{\chi\bm{p}+\bm{q}\downarrow}^{\dagger} z_{\chi\bm{p}\downarrow}] \nn && \psi^{\dagger}(\bm{p}+\bm{q}) \Bigl( \frac{1}{\sqrt{G_{\bm{p}+\bm{q}/2}}} \bm{A}_{\bm{q}} \cdot [\bm{v}_{\bm{p} + \bm{q}/2} + \bm{B}_{b} \bm{v}_{\bm{p} + \bm{q}/2} \cdot \bm{b}_{\bm{p}+\bm{q}/2}] \Bigr) \psi(\bm{p}) . \eqa Here, the solution of the semi-classical equation of motion has been introduced explicitly, where a constant background magnetic field $\bm{B}_{b}$ is applied for a Weyl metallic state but without electric field. The factor of $z_{\chi\bm{p}+\bm{q}\uparrow}^{\dagger} z_{\chi\bm{p}\uparrow} + z_{\chi\bm{p}+\bm{q}\downarrow}^{\dagger} z_{\chi\bm{p}\downarrow}$ counts the reduction of an overlap between wave functions at different momentum points, which originates from the spin-momentum locking. The gauge-matter coupling vertex becomes modified seriously according to the generalized Drude model for the chiral Fermi surface, where the chiral anomaly generates anomalous velocity terms. In the same way we are allowed to rewrite the Berry-phase term as follows \bqa && \mathcal{S}_{B} = - \int_{0}^{\beta} d \tau \int \frac{d^{3} \bm{p}}{(2\pi)^{3}} \sqrt{G_{\bm{p}}} \bm{a}_{\bm{p}} \cdot c_{\sigma}^{\dagger}(\bm{p}+\bm{q}) \dot{\bm{p}}_{\bm{p},\bm{q}} c_{\sigma}(\bm{p}) \nn && \longrightarrow - \int_{0}^{\beta} d \tau \int \frac{d^{3} \bm{p}}{(2\pi)^{3}} \sqrt{G_{\bm{p}}} \psi^{\dagger}(\bm{p}) \Bigl( \frac{1}{\sqrt{G_{\bm{p}}}} \bm{a}_{\bm{p}} \cdot (\bm{v}_{\bm{p}} \times \bm{B}_{b}) \Bigr) \psi(\bm{p}) . \eqa

The above discussion leads us to propose an effective field theory for a Weyl metallic phase
\bqa && Z_{W} = \int D \psi_{+}(\bm{p}) D \psi_{-}(\bm{p}) \exp\Bigl[- \int_{0}^{\beta} d \tau \int \frac{d^{3} \bm{p}}{(2\pi)^{3}} \sum_{\chi = \pm} \sqrt{G_{\bm{p}}^{\chi}} \Bigl\{ \psi_{\chi}^{\dagger}(\bm{p}) \Bigl( \partial_{\tau} - \mu \nn && + (|\bm{p}| - \bm{b}_{\bm{p}}^{\chi} \cdot \bm{B}) - \frac{1}{\sqrt{G_{\bm{p}}^{\chi}}} \bm{a}_{\bm{p}}^{\chi} \cdot (\bm{v}_{\bm{p}} \times \bm{B}_{b}) \Bigr) \psi_{\chi}(\bm{p}) + \int \frac{d^{3} \bm{q}}{(2\pi)^{3}} [z_{\chi\bm{p}+\bm{q}\uparrow}^{\dagger} z_{\chi\bm{p}\uparrow} + z_{\chi\bm{p}+\bm{q}\downarrow}^{\dagger} z_{\chi\bm{p}\downarrow}] \nn && \psi_{\chi}^{\dagger}(\bm{p}+\bm{q}) \Bigl( \frac{1}{\sqrt{G_{\bm{p}+\bm{q}/2}^{\chi}}} \bm{A}_{\bm{q}} \cdot [\bm{v}_{\bm{p} + \bm{q}/2} + \bm{B}_{b} \bm{v}_{\bm{p} + \bm{q}/2} \cdot \bm{b}_{\bm{p}+\bm{q}/2}^{\chi}] \Bigr) \psi_{\chi}(\bm{p}) \Bigr\} \Bigr] , \eqa where a pair of chiral Fermi surfaces has been marked explicitly with $\chi$. The linear dispersion of a Weyl electron on the chiral Fermi surface gets a correction from the Berry-phase term. The electric current coupled to the electromagnetic gauge field becomes also generalized due to the chiral anomaly.

\subsection{Enhancement of the longitudinal electrical-magneto conductivity within the Kubo formula}

One can derive the longitudinal positive electrical-magneto conductivity within the Kubo formula based on the effective field theory for a pair of chiral Fermi surfaces. Taking derivatives of the free energy $F_{W}[\bm{A}_{\bm{q}}] = - \frac{1}{\beta} \ln Z_{W}[\bm{A}_{\bm{q}}]$ twice with respect to the electromagnetic vector potential $\bm{A}_{\bm{q}}$, we obtain the following expression for the current-current correlation function
\bqa && \Pi_{ij}(\bm{q},i\Omega) = \sum_{\bm{p}} [z_{\chi\bm{p}+\bm{q}\uparrow}^{\dagger} z_{\chi\bm{p}\uparrow} + z_{\chi\bm{p}+\bm{q}\downarrow}^{\dagger} z_{\chi\bm{p}\downarrow}] [z_{\chi\bm{p}-\bm{q}\uparrow}^{\dagger} z_{\chi\bm{p}\uparrow} + z_{\chi\bm{p}-\bm{q}\downarrow}^{\dagger} z_{\chi\bm{p}\downarrow}] \nn && [\bm{v}_{\bm{p} + \bm{q}/2} + \bm{B}_{b} \bm{v}_{\bm{p} + \bm{q}/2} \cdot \bm{b}_{\bm{p}+\bm{q}/2}^{\chi}]_{i} [\bm{v}_{\bm{p} - \bm{q}/2} + \bm{B}_{b} \bm{v}_{\bm{p} - \bm{q}/2} \cdot \bm{b}_{\bm{p}-\bm{q}/2}^{\chi}]_{j} \nn && \frac{1}{\beta} \sum_{i\omega} G_{\chi}(\bm{p}+\bm{q},i\omega+i\Omega) G_{\chi}(\bm{p},i\omega) . \eqa Here, \bqa && G_{\chi}(\bm{p},i\omega) = \frac{1}{i \omega + \mu - (|\bm{p}| - \bm{b}_{\bm{p}}^{\chi} \cdot \bm{B}) + \frac{1}{\sqrt{G_{\bm{p}}^{\chi}}} \bm{a}_{\bm{p}}^{\chi} \cdot (\bm{v}_{\bm{p}} \times \bm{B}_{b}) + i \gamma_{imp} \mbox{sgn}(\omega)} \eqa is the propagator of the spinless fermion near the chiral Fermi surface, where the contribution of impurity scattering has been introduced within the Born approximation \cite{Review_Disorder}. $\gamma_{imp}$ is a scattering rate of the spinless quasiparticle.

Taking the limit of $\bm{q} \rightarrow 0$ in this polarization function, we obtain the dc conductivity
%
%\bqa && \sigma_{ij} = - \lim_{\Omega \rightarrow 0} \frac{\Im \Pi_{ij}(i\Omega \rightarrow \Omega + i \eta)}{\Omega}  \eqa
%
\bqa && \sigma_{ij} = \frac{1}{\pi} \sum_{\bm{p}} [\bm{v}_{\bm{p}} + \bm{B}_{b} \bm{v}_{\bm{p}} \cdot \bm{b}_{\bm{p}}^{\chi}]_{i} [\bm{v}_{\bm{p}} + \bm{B}_{b} \bm{v}_{\bm{p}} \cdot \bm{b}_{\bm{p}}^{\chi}]_{j} \nn && \int_{-\infty}^{\infty} d \varepsilon \Bigl( - \frac{\partial f(\varepsilon)}{\partial \varepsilon} \Bigr) \frac{\gamma_{imp}^{2}}{\Bigl\{\Bigl(\varepsilon + \mu - (|\bm{p}| - \bm{b}_{\bm{p}}^{\chi} \cdot \bm{B}) + \frac{1}{\sqrt{G_{\bm{p}}^{\chi}}} \bm{a}_{\bm{p}}^{\chi} \cdot (\bm{v}_{\bm{p}} \times \bm{B}_{b}) \Bigr)^{2} + \gamma_{imp}^{2}\Bigr\}^{2}} . \eqa
It is straightforward to perform the momentum integral. As a result, we recover the enhancement of the longitudinal electrical-magneto conductivity of the Boltzmann-equation approach with the topological Drude model \cite{Anomaly_CFS,NLMR_Exp,NLMR,Son_Boltzmann}
\bqa && \sigma_{xx} \approx \frac{1}{\pi} N_{F} v_{F}^{2} \int_{-1}^{1} d \cos \theta \Bigl(1 + \bm{B}_{b}^{x 2} |\bm{b}_{\bm{p}_{F}}^{\chi}|^{2} \cos^{2} \theta \Bigr) \int_{-\infty}^{\infty} d \xi \frac{\gamma_{imp}^{2}}{(\xi^{2} + \gamma_{imp}^{2})^{2}} \nn && = \frac{2}{\pi} N_{F} v_{F}^{2} \Bigl( \int_{-\infty}^{\infty} d x \frac{1}{(x^{2} + 1)^{2}} \Bigr) \Bigl(1 + \frac{1}{3} |\bm{b}_{\bm{p}_{F}}^{\chi}|^{2} \bm{B}_{b}^{x 2} \Bigr) \gamma_{imp}^{-1} \nn && \equiv \Bigl(1 + \frac{1}{3} [|\bm{b}_{\bm{p}_{F}}^{+}|^{2} + |\bm{b}_{\bm{p}_{F}}^{-}|^{2}] \bm{B}_{b}^{x 2} \Bigr) \sigma , \eqa where $\sigma$ is the residual conductivity determined by the impurity scattering rate $\gamma_{imp}$, the Fermi velocity $v_{F}$, and the density of states at the Fermi energy $N_{F}$.

\section{Summary and discussion}

In this review article, we have discussed two kinds of theoretical frameworks for anomalous transport phenomena in Weyl metal, where one is the Boltzmann-equation approach and the other is an effective field-theory approach. An essential point was how to encode the topological structure of the Weyl metallic state, given by both the Berry curvature and chiral anomaly. In both theoretical frameworks the idea was to generalize the Drude model for dynamics of the Weyl electron on the chiral Fermi surface. Resorting to the topological Drude model, we could construct not only the Boltzmann-equation approach but also a topological Fermi-liquid field theory for the topological metallic phase with a pair of chiral Fermi surfaces. Both approaches gave rise to the enhancement of the longitudinal electrical-magneto conductivity consistently.

\subsection{Derivation of the Boltzmann equation with the topological Drude model}

We would like to emphasize that the present phenomenological Boltzmann-equation approach is applicable only when the chemical potential lies away from the Weyl point, forming a pair of chiral Fermi surfaces. When the chemical potential touches the Weyl point, we should re-derive the Boltzmann equation from the effective field theory for quantum electrodynamics in one-time and three-space dimensions (QED$_{4}$) under the magnetic field. The distribution function in this relativistic Boltzmann equation will be expressed as a $4 \times 4$ matrix since the lesser Green's function consists of the four-component spinor. An interesting and fundamental problem is as follows. Taking the non-relativistic limit from the matrix Boltzmann equation when the chemical potential lies above the Weyl point, can we reproduce the present Boltzmann-equation framework, where effects of other components except for the Fermi-surface component are ``integrated out" or ``coarse grained", giving rise to such contributions as semi-classical equations of motion? This research will give a formal basis to the present phenomenological Boltzmann-equation approach. Recently, the Boltzmann-equation framework has been derived from QED$_{4}$, based on the introduction of the Wigner function to satisfy a quantum kinetic equation \cite{QBE_WM3,QBE_WM1,QBE_WM2}. These derivations imply that Lorentz symmetry, gauge symmetry, and quantum mechanics are important ingredients for the existence of chiral anomaly.

\subsection{Derivation of the topological Fermi-liquid theory}

We would like to point out that the standard diagrammatic approach based on QED$_{4}$ fails to incorporate contributions from chiral anomaly for some physical quantities. Although the diagrammatic approach based on QED$_{4}$ with a finite chemical potential succeeds in calculating the anomalous Hall effect \cite{AHE_Kubo1,AHE_Kubo2} and chiral magnetic effect \cite{CME_Kubo1,CME_Kubo2}, both of which are associated with transverse contributions of the current-current correlation function, it fails to describe the negative longitudinal magneto-resistivity, associated with the longitudinal component of the current-current correlation function. An effective action for single particle dynamics has been constructed in the phase pace, imposing the momentum-space Berry gauge-field. Reformulating the Berry phase based on the Stoke's theorem, a five-dimensional gauged Wess-Zumino-Witten term has been proposed to reproduce chiral anomalies associated with not only electromagnetic but also gravitational fields and their mixing \cite{WM_WZW}. If this action functional is reformulated in the second quantized form, an effective field theory for the topological Fermi-liquid state will be given by the Landau's Fermi-liquid theory backup with the five dimensional gauged Wess-Zumino-Witten term. Recall the problem of a spin chain, where the spin$-1/2$ chain is described by the O(3) nonlinear$-\sigma$ model backup with a Berry phase term while such a topological term is irrelevant for the spin$-1$ chain with a periodic boundary condition \cite{Spin_Textbook}. Unfortunately, this fascinating construction does not seem to be much practical for actual calculations because it is not easy to introduce the role of such a term into diagrammatic analysis as the case of field theories with topological terms.

On the other hand, we have introduced the role of both the Berry curvature and chiral anomaly within the U(1) slave-fermion representation. The orthogonality of the wave-function overlap and the Berry gauge field in momentum space are expressed in terms of the CP$^{1}$ bosonic field while dynamics of the pair of chiral Fermi surfaces is described by the spinless fermion field, where the chiral anomaly plays the role of an effective potential, modifying the dispersion of the Weyl electron and the gauge coupling with the spinless fermion field. This phenomenological construction should be derived from QED$_{4}$ under magnetic fields, where high energy fluctuations except for low energy modes near the Fermi surface are ``integrated out" or ``coarse grained", expected to cause anomaly-induced effective potentials as suggested in the present manuscript \cite{EFT}. We believe that this way serves a novel direction, deriving the chiral anomaly in the presence of a Fermi surface.

\subsection{Thermal transport and violation of the Wiedemann-Franz law}

Recently, one of the authors has investigated thermal transport coefficients based on the Boltzmann transport theory with the topological Drude model \cite{WF_Law_WM}. We found that not only the longitudinal electrical-magneto conductivity but also both the Seebeck and thermal conductivities in the longitudinal setup show essentially the same enhancement proportional to $B^{2}$, the origin of which is the chiral anomaly. Actually, our preliminary experimental result on the longitudinal thermal conductivity seems to show the similar behavior as the longitudinal electrical-magneto conductivity \cite{Thermal_WM_Exp}. An interesting prediction of this theoretical study is that the Wiedemann-Franz law will be violated only in the longitudinal setup, showing the $B^{2}$ dependence in the Lorentz number. On the other hand, the Wiedemann-Franz law turns out to hold in the transverse setup as expected. This prediction is quite surprising because the concept of an electron quasiparticle is still valid. The origin of the violation is that the coefficient in front of $B^{2}$ differs from each transport coefficient, i.e., $\mathcal{C}_{ABJ}^{el} \not= \mathcal{C}_{ABJ}^{th}$, where the superscript of $el$ ($th$) means electrical (thermal). Since the breakdown of the Wiedemann-Franz law appears from this term, the chiral anomaly is the mechanism, implying that this metallic state should be distinguished from the Landau's Fermi-liquid state.

\subsection{The connection between the chiral anomaly and the surface Fermi-arc state}

In this paper we have focused on the bulk transport since we considered a metallic phase with a pair of Fermi surfaces. However, the emergence of the surface Fermi-arc state is regarded as the hallmark of the Weyl metallic state \cite{Weyl_Metal3}, exactly analogous to the Weyl point on the surface state of a topological insulator, where each Fermi point of the Fermi arc corresponds to the Weyl point of the case of the topological insulator. The standard argument for the existence of the surface Fermi-arc state is as follows \cite{Weyl_Metal3}. Suppose a pair of Weyl points in the $z-$direction, given by  $(0,0,- g_{\psi} B / v)$ and $(0,0, g_{\psi} B / v)$, respectively. If we fix the momentum of $\bm{k}_{z}$, we are in the two dimensional plane, allowing us to define the Chern number which counts the number of magnetic monopoles in momentum space. Considering the two dimensional plane at $(0,0,k_{z})$ with $k_{z} < - g_{\psi} B / v$, we see that the Chern number vanishes. However, considering two dimensional planes with $- g_{\psi} B / v < k_{z} < g_{\psi} B / v$, we find that the Chern number is one, since such two dimensional planes cross the Weyl point. If the plane lies in $g_{\psi} B / v < k_{z}$, it crosses both the $+$ and $-$ charges, causing the Chern number to vanish. Since the two dimensional planes with the Chern number $1$ are essentially the same as the integer quantum Hall state, we have an edge state for each $k_{z}$ within $- g_{\psi} B / v < k_{z} < g_{\psi} B / v$, giving rise to a Fermi arc state on the surface. Unfortunately, this spectroscopic fingerprint has not been observed yet.

In our opinion the connection between the chiral anomaly and the surface Fermi-arc state has not been clarified yet. Although the Hall current through the bulk channel can be derived from the topological-in-origin $\bm{E} \cdot \bm{B}$ term \cite{Weyl_Metal3}, the connection between the bulk Hall current and surface Hall current is not discussed completely as far as we understand. We believe that this issue should be more clarified.

\section*{Acknowledgement}

This study was supported by the Ministry of Education, Science, and Technology (No. 2012R1A1B3000550 and No. 2011-0030785) of the National Research Foundation of Korea (NRF) and by TJ Park Science Fellowship of the POSCO TJ Park Foundation. HJ was supported by Basic Science Research Program through the National Research Foundation of Korea (NRF) funded by the Ministry of Education, Science, and Technology (No. 2014R1A1A1002263). MS was supported by YO-COE Foundation from Yamagata University. KS appreciates enlightening discussions with Suk-Bum Chung and Sang-Jin Sin.

\appendix

\section{What makes Weyl metal special: Axion electrodynamics}

The effective Dirac Hamiltonian of $H_{eff} = \int \frac{d^{3} \boldsymbol{k}}{(2\pi)^{3}} \psi_{\sigma\tau}^{\dagger}(\boldsymbol{k}) \Bigl( v \boldsymbol{k} \cdot \boldsymbol{\sigma}_{\sigma\sigma'} \otimes \boldsymbol{\tau}_{\tau\tau'}^{z} + m(|\boldsymbol{k}|) \boldsymbol{I}_{\sigma\sigma'} \otimes \boldsymbol{\tau}_{\tau\tau'}^{x} \Bigr) \psi_{\sigma'\tau'}(\boldsymbol{k})$ in momentum space gives its corresponding effective action of $\mathcal{S}_{eff} = \int_{0}^{\beta} d \tau \int d^{3} \bm{r} \psi_{\sigma\tau}^{\dagger}(\boldsymbol{r},\tau) \left\{ \partial_{\tau} \boldsymbol{I}_{\sigma\sigma'} \otimes \boldsymbol{I}_{\tau\tau'} - i v \bm{\nabla} \cdot \boldsymbol{\sigma}_{\sigma\sigma'} \otimes \boldsymbol{\tau}_{\tau\tau'}^{z} + m \boldsymbol{I}_{\sigma\sigma'} \otimes \boldsymbol{\tau}_{\tau\tau'}^{x} \right\} \psi_{\sigma'\tau'}(\boldsymbol{r},\tau)$ in real space. Representing two kinds of Pauli matrices in terms of Dirac gamma matrices of $\gamma^0 = \boldsymbol{I}_{\sigma\sigma'} \otimes \boldsymbol{\tau}_{\tau\tau'}^{x}$ and $\gamma^{k} = - i \boldsymbol{\sigma}_{\sigma\sigma'}^{k} \otimes \boldsymbol{\tau}_{\tau\tau'}^{y}$, we are allowed to rewrite this expression as the standard representation of the Dirac theory, given by $\mathcal{S}_{eff} = \int d^{4} x \bar{\psi}(x) \left\{i \gamma_{0} \partial_{\tau} - i v \boldsymbol{\gamma} \cdot \boldsymbol{\nabla} + m \right\} \psi(x)$, where $\bar{\psi}(x) = \psi^{\dagger}(x) \gamma^{0}$ is a conjugate partner of $\psi(x)$.

Introducing electromagnetic fields into this expression, we obtain QED$_{4}$ (quantum electrodynamics in one time and three spatial dimensions) \bqa && \mathcal{S}_{QED} = \int d^{4} x \Bigl\{ \bar{\psi}(x) \Bigl( i \gamma^{\mu} [\partial_{\mu} + i e A_{\mu}] + m \Bigr) \psi(x) - \frac{1}{4} F_{\mu\nu} F^{\mu\nu} + \theta(x) \frac{e^{2}}{16 \pi^{2}} \epsilon^{\mu\nu\rho\delta} F_{\mu\nu} F_{\rho\delta} \Bigr\} , \label{QED4} \nn \eqa where the topological-in-origin $\bm{E}\cdot\bm{B}$ term has been incorporated. $F_{\mu\nu} = \partial_{\mu} A_{\nu} - \partial_{\nu} A_{\mu}$ is the field strength tensor and $\mathcal{L}_{EM} = - \frac{1}{4} F_{\mu\nu} F^{\mu\nu} = \frac{1}{2} (\bm{E}^{2} - \bm{B}^{2})$ recovers Maxwell equations through the variational principle. $\epsilon^{\mu\nu\rho\delta}$ is an antisymmetric tensor and $\epsilon^{\mu\nu\rho\delta} F_{\mu\nu} F_{\rho\delta}$ is nothing but the $\bm{E} \cdot \bm{B}$ term.

Replacing the topological-in-origin term with a chiral current based on the anomaly equation \bqa && \partial_{\mu} (\bar{\psi} \gamma^{\mu} \gamma^{5} \psi) = - \frac{e^{2}}{16 \pi^{2}} \epsilon^{\mu\nu\rho\delta} F_{\mu\nu} F_{\rho\delta} , \eqa one may rewrite the above expression as follows \bqa && \mathcal{S}_{WM} = \int d^{4} x \Bigl\{ \bar{\psi}(x) \Bigl( i \gamma^{\mu} [\partial_{\mu} + i e A_{\mu}] + m + c_{\mu} \gamma^{\mu} \gamma^{5} \Bigr) \psi(x) - \frac{1}{4} F_{\mu\nu} F^{\mu\nu} \Bigr\} , \label{WM_Action} \nn \eqa where the chiral gauge field $c_{\mu} = (c_{\tau}, \bm{c})$ appears through the technique of the integration by part, given by $c_{\mu} = \partial_{\mu} \theta(x)$.

It is quite interesting to observe that the Zeeman energy term plays essentially the same role as the chiral gauge field, given by \bqa H_{TRB} = g_{\psi} \psi_{\sigma\tau}^{\dagger}(\boldsymbol{k}) (\boldsymbol{B} \cdot \boldsymbol{\sigma}_{\sigma\sigma'} \otimes \boldsymbol{I}_{\tau\tau'}) \psi_{\sigma'\tau'}(\boldsymbol{k}) \equiv \bar{\psi} (\boldsymbol{c} \cdot \boldsymbol{\gamma} \gamma_{5})\psi , \eqa where the chiral gauge field is identified with $\bm{c} = g_{\psi} \bm{B}$ and $c_{\tau} = 0$. This indicates that the axion angle is given by $\theta(\bm{r}) = g_{\psi} \bm{B} \cdot \bm{r}$.

It is straightforward to integrate over gapped fermion excitations, resulting in an effective field theory for electromagnetic fields \bqa && \mathcal{L}_{axion} = - \frac{1}{4} F_{\mu\nu} F^{\mu\nu} + \theta(\bm{r},t) \frac{e^{2}}{16 \pi^{2}} \epsilon^{\mu\nu\rho\delta} F_{\mu\nu} F_{\rho\delta} , \eqa where time dependence in $\theta(\bm{r},t)$ has been introduced for generality. Applying the least action principle to this effective Lagrangian, we reach Maxwell equations to describe the axion electrodynamics \bqa && \bm{\nabla} \cdot \bm{D} = 4 \pi \rho + 2 \alpha (\bm{\nabla} P_{3} \cdot \bm{B})  , \nn && \bm{\nabla} \times \bm{H} - \frac{1}{c} \frac{\partial \bm{D}}{\partial t} = \frac{4\pi}{c} \bm{j} - 2 \alpha \Bigl( (\bm{\nabla} P_{3} \times \bm{E}) + \frac{1}{c} (\partial_{t} P_{3}) \bm{B} \Bigr) , \nn && \bm{\nabla} \times \bm{E} + \frac{1}{c} \frac{\partial \bm{B}}{\partial t} = 0 , ~~~~~ \bm{\nabla} \cdot \bm{B} = 0 , \eqa where we follow the standard cgs notation with $P_{3}(\bm{r},t) \propto \theta(\bm{r},t)$ and the fine structure constant $\alpha$ \cite{Axion_EM}.

\end{document}